\documentclass[aps,showpacs,twocolumn,showkeys,amsmath,amssymb]{revtex4}
\usepackage{graphicx}
\usepackage{dcolumn}
\usepackage{bm}
\usepackage{mathrsfs}
\usepackage{epsfig}

\newcommand{\N}{N_c}

\begin{document}

\title{The total  nucleon-nucleon cross section at large $N_c$}

\author{Thomas D. Cohen}
\email{cohen@physics.umd.edu}
\affiliation{Department of Physics, University of Maryland, College Park, MD 20742-4111, USA}

\date{\today}

\begin{abstract}

It is shown that at sufficiently  large $N_c$    for incident momenta which are much larger than the  QCD, the total nucleon-nucleon cross section is independent of incident momentum and   given by $\sigma^{\rm total}=2  \pi   \log^2(N_c) / (  m^2_{\pi}) $ .  This result is valid in the extreme large $N_c$ regime  of $\log(N_c)  \gg 1$ and has corrections of relative order  $\log \left( \log(N_c) \right )/\log(N_c)$.   A possible connection of this result to the Froissart-Martin bound is discussed.
\end{abstract}

\maketitle

The large $N_c$ limit of QCD and the $1/N_c$ expansion have been of great interest since introduced by `tHooft nearly 40 years ago \cite{tHooft}.  While the approach to date has not provided a path by which quantities can be calculated {\it ab initio} directly  from QCD except in special cases such as QCD in 1+1 dimensions \cite{tHooft2} or QCD in the limit of heavy quark masses \cite{Witten,CohenKumarNdousse}, it  has provided a qualitative understanding of many aspects of hadronic phenomena.  Witten's extension  of the analysis to include baryons has played a critical role \cite{Witten}.  One of the remarkable features of baryons is the emergence of a contract $SU(2N_f)$ symmetry at large $N_c$  \cite{SU2Nf}   which has allowed predictions  of both ground state baryons \cite{SU2Nf}  and excited baryonic resonance  \cite{excited}.  In this work we will focus on the nonstrange sector and assume exact isospin invariance.

Implications of large $N_c$ QCD for nuclear physics were first explored in Witten's seminal paper on large $N_c$ baryons \cite{Witten}.  A key result of this analysis is that the nucleon-nucleon interaction has a strength which scales as $N_c^1$ and a range which scales as $N_c^0$.  Moreover,  nucleon-nucleon scattering with fixed incident  momentum has no smooth large $N_c$ limit.  However, a sensible time-dependent mean-field description emerges if the initial velocity is held fixed at large $N_c$ (that is, that momentum scales linearly with $N_c$ given that the mass is linear in $N_c$).  While there has been significant work on various aspects of nuclear physics at large $N_c$,  such as treatments of the NN potential \cite{NNpotential}, the phenomenological relevance of the large $N_c$ limit for nuclear physics is far less clear than for hadronic physics \cite{disc}.  Despite this, it is of interest to understand the $N_c$ scaling behavior of quantities of interest in nuclear physics. One quantity which has received comparatively little attention except for a recent paper on its spin flavor dependence \cite{CohenGelman}  is the total nucleon-nucleon cross section (with the effects of electromagnetic interactions removed).   This is unfortunate since,  as will be shown in this letter, the total cross section is  truly remarkable in that it can be computed analytically when $N_c$ is sufficiently large:
\begin{equation}
 \sigma^{\rm total}=\frac{2  \pi  \log^2(N_c) } { m^2_{\pi}}  \; . \label{result}
 \end{equation}
 Equation~(\ref{result}) holds for  all spin-isopsin channels in the regime where the incident momentum is much larger than $\Lambda_{\rm QCD}$;  corrections to  Eq.~(\ref{result}) are of relative order $\log \left( \log(N_c) \right )/\log(N_c)$.   Formally  $N_c$ needs to be extremely large for Eq.~(\ref{result}) to  hold and it is not obvious that the result is phenomenologically  relevant for the physical world of $N_c=3$.  In any event,  the result is of real interest  from the perspective of theory.

To gain insight, it is useful to first consider a simplified ``toy problem'' of scattering of  two nonrelativistic  spinless particles of  mass $M$ interacting via a central potential  which falls off exponentially at large distances.   The problem has a control parameter, $\lambda$, which controls the scaling of the potential strength (but not its range), the mass, and the initial relative momentum:
\begin{equation}
M=\lambda \tilde{M} \; \; \;
V(r) = \lambda \tilde{V}(r)\; \; \;
k = \lambda \tilde{k} \, ; \label{scale}
\end{equation}
 where the quantities with a tilde are independent of $\lambda$.
The classical trajectory followed by a particle in this problem depends on the impact parameter $b$ and $\tilde{p}$, $\tilde{V}(r)$, and $\tilde{M}$ but {\it not} on $\lambda$: if  the quantum scattering is  described well classically, then the differential cross section will be independent of $\lambda$.   The parameter, $\lambda$, however, controls the region of validity of a semi-classical description in an underlying  quantum scattering problem;  the classical limit corresponds to large $\lambda$.  Of course,  by design this problem  mirrors the $N_c$  scaling rules of NN scattering with $\lambda$ playing the role of $N_c$.

The potential is central and the partial waves are independent.
Using the scaling rules in Eq.~(\ref{scale}) and  the Schr\"odinger equation for a given partial wave,   the phase shifts can be shown to scale as
 \begin{equation}
 \delta_l(k) = \delta_{\lambda \tilde{l}}(\lambda \tilde{k}) =\lambda \,  \tilde{\delta}_{\tilde{l}}(\tilde{k}) \; ,
 \label{psscale}\end{equation}
 where $\tilde{l}=l/\lambda$ is introduced for convenience and $\tilde{\delta}$ is independent of $\lambda$.  Corrections are of relative order $1/\lambda$.  To see this, parameterize  the radial wave function for a given partial wave as the product of a phase and an amplitude, $\psi_l(r)=e^{i \Phi_l(r)}|\psi_l(r)|$, and take $l$ to be proportional to $\lambda$.  It is easy to show self-consistently that at large $\lambda$, $\Phi$ is proportional to $\lambda$ while $  |\psi_l(r)|$ is slowly varying and independent of $\lambda$.   This is precisely what one expects if $\lambda$ acts as the control  parameter for the semi-classical limit.
The total cross section for central potentials is given by
 \begin{equation}
 \begin{split}
 \sigma^{\rm total}_{\rm toy}(k)&=\frac{4 \pi}{k^2}  \sum_l (2l+1) \sin^2(\delta_l(k))  \\
 &\approx  \frac{4 \pi}{\tilde{k}^2}  \int d \tilde{l} \, 2\tilde{l} \sin^2(\lambda \tilde{\delta}_{\tilde{l}} (\tilde{k}) )
\end{split}
\label{pw} \end{equation}
 where $\delta_l$ is the phase shift for the $l^{\rm th}$ partial wave and the integral expression becomes exact as $\lambda \rightarrow \infty$.

In the integral of Eq.~(\ref{pw}),  focus on the range from $\tilde{l}_1$ to  $\tilde{l}_2 $ .  For large $\lambda$,  $\sin^2$  oscillates rapidly and averages to $\frac{1}{2}$ (up to corrections of order $1/\lambda$)  over this region; the contribution to the  cross section becomes
\begin{equation}
 \Delta \sigma^{\rm total}_{\rm toy}(k) = 2\pi (b_2^2 - b_1^2) \; \; {\rm with } \; \; b \equiv \frac{l}{k}=\frac{\tilde{l}}{\tilde{k}} \; .
\label{Delta} \end{equation}
This is twice the geometric cross section associated with impact parameters from $b_1$ and $b_2$; the factor of 2 is due to a nearly forward diffractive scattering contribution equal to the geometrical contribution \cite{Jac}.  At infinite $\lambda$  there is no bound on the $\tilde{l}$s  that contribute implying that the total cross section diverges as $\lambda  \rightarrow \infty$.

 The  quantum cross section is finite because  the phase shifts approach zero as $l \rightarrow \infty$.  For any  finite value of  $\lambda$, there is a regime of  sufficiently large $l$  such that $\tilde{\delta} \sim \lambda^{-1}$ and  the $\sin^2$ term does not oscillate rapidly to yield an average of $\frac{1}{2}$.  The phase shift in this regime rapidly becomes  small and  makes small contributions to the total cross section.  The value of $\tilde{l}$ beyond which  the rapid oscillations effectively turns off  depends on $\lambda$ and gets pushed off to infinity as $\lambda \rightarrow \infty$.


 To see this, start with the integral form of Eq.~(\ref{pw}) and change variables into an integral over $\tilde{\delta}$.  It is a simple matter to show that
 \begin{equation}
 \sigma^{\rm total}_{\rm toy}=  2\pi b_{\rm cut}^2  +  2 \pi \tilde{k}^{-2} \int_0^\frac{\delta_{\rm cut}}{\lambda}  d \tilde{\delta}  \left ( \frac {d \tilde{l}^2}{ d \tilde{\delta} } \right ) \,  \sin^2(\lambda \tilde{\delta})   + {\cal O} (\lambda^{-1})
\label{cut}\end{equation}
where $\frac {d \tilde{l}^2}{ d \tilde{\delta} } $ is treated as a function of $\tilde{\delta}$ in the integral and $\delta_{\rm cut}$ is an arbitrary but fixed ``cutoff'' phase shift of order $\lambda^0$; $b_{\rm cut}$ is the impact parameter associated with  $\delta_{\rm cut}$.  The arbitrariness in the choice of  $\delta_{\rm cut}$ is compensated by the integral in Eq.~(\ref{cut}).  If the integral  in Eq.~(\ref{cut}) is parametrically smaller than  $2 \pi b_{\rm cut}^2$ when $\delta_{\rm cut}$ is of order $\lambda^0$, then up to parametrically small corrections $\sigma^{\rm total}_{\rm toy}=2 \pi b_{\rm cut}^2$.

As shall be shown self-consistently, the total cross section is dominated by the behavior for large impact parameters, or equivalently, large  $\tilde{l}$.  In this regime, each partial wave is semi-classical and the potential is much smaller than the centrifugal barrier.  Thus,  the phase shifts are well approximated \cite{L&L}   by
 \begin{equation}
\tilde{ \delta} = - \int_{\tilde{l}/\tilde{k}}^\infty  \frac{\tilde{\mu}\tilde{V}(r)}{\sqrt{\tilde{k}^2- \tilde{l}^2/r^2} } dr \, .
\label{sc}\end{equation}
where $\tilde{\mu}$ is the reduced mass divided by $\lambda$.

For concreteness, take the form of the potential to be  the sum of Yukawa interactions:  $\tilde{V}(r) = \sum_n \tilde{C}_n \frac{\exp({-r/r_n})}{r} $ where the $\tilde{C}_n$  are strength parameters  independent of $\lambda$ and the $r_n$ are the ranges.  Note that at large $\lambda$, $b_{\rm cut}$ is large and the integral in Eq.~(\ref{sc}) is dominated by the longest-range contribution to the potential.  Evaluating the integral  yields
\begin{equation}
\tilde{ \delta}_{\tilde{l}} =-\frac{ \tilde{C} \tilde{M}}{\tilde{k}} {K_0} \left(\tilde{l}/(\tilde{k} r_0) \right )=-\frac{ \tilde{C_0} \tilde{\mu}}{\tilde{k}}  { K_0}\left( b_{\tilde{l}}/ r_0) \right ) \; .
 \end{equation}
 where $r_0$ is the longest range in the potential and $\tilde{C}_0$ is the associated strength.
For large values of $ b_{\tilde{l}}/ r_0$ it is legitimate to use the asymptotic form of the Bessel function when inverting this relation; doing this yields
 \begin{equation}
 b_{\tilde{l}}= \frac{r_0}{2} \, W \left( \frac{\tilde{C}^2 \tilde{\mu}^2 \pi }{ \tilde{\delta}_{\tilde{l}}^2 \tilde{k}^2  }
\right )
\end{equation}
where $W$ is the Lambert function.  As $x$ gets very large  $W(x) \rightarrow \log (x) $ (reflecting the dominantly  exponential behavior of  $K_0$)   with corrections of relative order  $ \log\left( \log(x) \right )/\log(x)$.  At  large ${\tilde{l}}$, the phase shifts become small;  the log is dominated by $\tilde{\delta}_{\tilde{l}}$;  $b_l \approx  -r_0 \log(\tilde{\delta_l})$ and  $b_{\rm cut} =- r_0 \log(\delta_{\rm cut}/\lambda)$.  Thus up to corrections of order $\lambda^0$
\begin{equation}
b_{\rm cut} =r_0 \log({\lambda})
\label{bcut}
\end{equation}
and  at large $\lambda$ is $ \sigma^{\rm total}_{\rm toy}=2  \pi  \, r_0^2 \,   \log^2(\lambda) $    provided the  integral in   Eq.~(\ref{cut}) is parametrically small---which, as will  be shown shortly, it is.

Note that  the  sensitivity to the particle's mass and to the strength of the potential  are contained in the order  $\lambda^0$ correction terms to Eq.~(\ref{bcut}). Note further,  that the sensitivity to the choice of $\delta_{\rm cut}$ is also contained in the $\lambda^0$ correction terms.  Since the dependence of the choice of $\delta_{\rm cut}$ is compensated by the integral in   Eq.~(\ref{cut}), it follows that the integral  is also  parametrically of order $\lambda^0$ and  makes a negligible contribution to the cross section at large $\lambda$.      Thus,
$ \sigma^{\rm total}_{\rm toy}=2  \pi  \, r_0^2 \,   \log^2(\lambda) $ with corrections  of relative order  $  \log\left( \log(x) \right )/\log(x) $.    With the substitutions  $r_0 \rightarrow  1/m_\pi$ and $\lambda \rightarrow  N_c$,  this is of the form of Eq.~(\ref{result}).  It should be apparent that  any power law prefactor to the Yukawa potentials cannot alter this result at leading order.

The result also holds for a relativistic version of the toy problem.  Consider  potential scattering in a relativistic two-body model  (which lacks  micro causality but can be consistently formulated as a quantum theory  \cite{rel}).   The basic set up remains intact: the partial wave decomposition still holds  and Eqs.~(\ref{pw}) and (\ref{cut}) remain valid.  In the semi-classical regime with sufficiently large $\tilde{l}$,   one  can always cast the phase shift into  the form of Eq.~(\ref{sc}) with $\tilde{V}(r)$ replaced by  (an energy dependent)  $\tilde{V}_{\rm eff}(r)$ whose  form depends on the transformation properties of the interaction.  If  the longest range interaction in the model transforms  as a  Lorentz scalar (as in QCD), then at long range $\tilde{V}_{\rm eff}(r) =\tilde{V}_s(r) \tilde{E}/\tilde{\mu}$. Relativity affects $b_{\rm cut}$ only by renormalizing the strength of the longest-range interaction by an 
energy-dependent amount independent of $\lambda$.  Since the leading behavior of the total cross section does not depend on the strength,  the relativistic toy model also has: $ \sigma_{\rm rel. toy}^{\rm total}=2  \pi  \, r_0^2 \,   \log^2(\lambda)  $ which corresponds to Eq.~(\ref{result}).

 Nucleon-nucleon scattering in large $N_c$ QCD is clearly more complicated than in  the toy problem for several reasons:  i) The nucleons have spin and the partial wave expansion for elastic scattering  is necessarily of a coupled channel form; ii) there is an emergent spin-isospin symmetry at large $N_c$ \cite{SU2Nf};  as a result of this symmetry the $\Delta$  and a whole tower of baryons are stable and  nearly  degenerate with the nucleon at large $N_c$.  The emergent symmetry implies correlations between channels in scattering \cite{NNpotential,CohenGelman,NNobservables}; iii) There are inelastic channels due to meson production.   However, as discussed below, the result in  Eq.~(\ref{result}) is quite robust and is unaltered by these complications.

To treat the  full problem, the generically strong (order $N_c$) nature of the NN interaction must be encoded in a model-independent way.  The potential model treatments of the toy problem are inappropriate to this problem and, in any event, the potential is intrinsically unphysical which can  lead to subtleties in the $N_c$ counting \cite{subtlty}.  The physically relevant object is the S-matrix for nucleon-nucleon scattering, ${\bf S}^{\rm NN}$.  Its elements for elastic scattering can be denoted $S^{\rm NN}_{l,a:l' ,a' }$ where $a$ ($a'$) specifies the spin and isospin configuration of the incident (final) states of the two nucleons.   Conservation of angular momentum and isospin along with the fermion nature of nucleons constrain the form of $ {\bf S}^{\rm NN}$; for example, the matrix elements are zero unless $l'=l-1,l,l+1$.

 The strong NN interaction at large $N_c$  cannot be encoded by having $ {\bf S}^{\rm NN}$  scale linearly in $N_c$; it is bounded due to the unitarity.  To proceed,  note  that in the toy problem,  Eq.~(\ref{psscale})  the phase shift, {\it i.e} the logarithm of the S matrix in the partial wave channel,  scales with $N_c$.  As will be discussed in detail in a future publication, this  behavior should hold generically for diagonal matrix elements of the S matrix in large $N_c$ QCD.
\begin{equation}
\log \left ( S_{l,a;la}^{\rm NN} \right ) \equiv 2 i \delta_{l, a}= 2 i \delta_{l, a}^R - 2 \delta_{l, a}^I  \sim N_c
\label{generic}\end{equation}
where the S-matrix is  for nucleon-nucleon elastic scattering at fixed initial velocity.  Note that $ \delta_{l, a}$ is not  real in general; the imaginary part reflects scattering out of the original channel either to other elastic channels (with different final $l$ or $a$) or to inelastic channels.  Both the real and imaginary parts are expected to  scale with $N_c$.  This applies to all physical channels ({\it eg.}, two neutrons with spins aligned with the beam).

The total cross section in multi-channel problems with the initial nucleons in spin-isospin configuration $a$ can be expressed in terms of the diagonal matrix elements of the S matrix:
\begin{align}
&\sigma^{\rm total}_a  (k) =\label{pw2}\\
& \frac{2 \pi}{k^2}  \sum_l (2l+1)  \left (1- \exp(-2 \delta^I_{l,a}(k)) \cos(2 \delta^R_{l,a}(k)) \right ) \approx  \nonumber\\
 &\ \frac{4 \pi}{\tilde{k}^2}  \int d \tilde{l}^2  \left (1- \exp \left (-2 N_c \delta^I_{\tilde{l},a}(N_c \tilde{k}) \right ) \cos \left (2N_c  \delta^R_{\tilde{l},a}(N_c \tilde{k})\right ) \right ) \nonumber
 \end{align}
where $k=N_c \tilde{k}$ ;  the first form is general  \cite{L&L} and the second form builds in the $N_c$ scaling of the phase shifts.  The integral form becomes exact in the limit $N_c \rightarrow \infty$.

Note that Eq.~(\ref{pw2}) coincides with  Eq.~(\ref{pw}) of the toy problem if one sets  $\delta^I_{l,a}$ to zero. Moreover, Eq.~(\ref{Delta}) continues to hold providing the integrand is in the regime where either   the real or imaginary parts of $\tilde{\delta}$ (or both) are of order unity; in the case of the real part it is due to rapid oscillations as in the toy problem; in the case of the imaginary part it holds due to an exponential suppression.
As in the toy problem, the total cross section is determined by  where $\tilde{\delta}$ ceases to be of order unity and becomes of order $1/N_c$.  It is easy to see that if, as a function of $l$, $\delta^I_{l,a}$  approaches zero at least as rapidly as $\delta^R_{l,a}$, then the total cross section at leading order will be determined by where $\tilde{\delta}^R_{\tilde{l},a}$ drops to order of $1/N_c$.

It is clear that  $\delta^I_{l,a}$  {\it does} approaches zero more rapidly than $\delta^R_{l,a}$: consider going to sufficiently large $\tilde{l}$ so that the phase shifts are accurately described by the Born approximation for the longest range part of the interaction---one-pion exchange.  In that regime the   $\delta^R_{l,a}$ is small but non-vanishing while $\delta^I_{l,a}$ vanishes at the first Born approximation level only arising at second order.   Moreover, it is clear that at very large $l$ the real part of the  phase shift is dominated by the Born approximation contribution to one-pion exchange  which drops off exponentially in exactly the same way that it does in  the toy model.  Accordingly the result in the toy model carries across and Eq.~(\ref{result}) follows exactly as in the toy model.  Note that this exponential fall-off holds for any physical spin and isospin configuration of the initial baryons; pion exchange dominates regardless of the initial spin orientations or whether the two nucleons are the same or different.    Thus  Eq.~(\ref{result}) also holds for any initial configuration and the leading order cross section is spin and isospin independent.  This is consistent with the analysis of Ref.~ \cite{CohenGelman}, although it is more restrictive than the most general leading-order result deduced there.

It is worth observing that at $N_c$, elastic scattering should account for 1/2 of the total scattering or more.  In the regime of $l$ where   $\delta^I_{l,a} \sim N_c$, the contribution to the scattering looks like a black disk for which diffractive scattering is 50\%.  There may, in principle, also be substantial contributions for the regime where $\delta^I_{l,a}$ is small but  $\delta^R_{l,a} \sim N_c$; such contributions will have elastic contributions of greater than 50\%.

The $\log^2$ form of the cross section in Eq.~(\ref{result}) is strikingly similar to the Froissart-Martin bound \cite{FM}.  At large Mandelstam $s$, considerations of unitarity, analyticity plus the knowledge that the pion is the lightest excitation in the system serve to bound the growth of the total cross section at large $s$
 \begin{equation}
\sigma^{\rm total} \le \frac{\pi}{m_\pi^2} \log^2 \left( \frac{s}{s_0} \right) \; ,
\label{FM}
\end{equation}
where $s_0$ is a reference scale.  This similarity  may not be accidental.  Note that  the natural regime for nucleon-nucleon scattering at large $N_c$ is for fixed velocity \cite{Witten} which in turn implies that $s \sim N_c^2$. Provided that $s_0$ does not also scale with $N_c$, the bound becomes $\sigma^{\rm total} \le \frac{\pi}{m_\pi^2} \log^2 \left( \frac{N_c \tilde{s}}{s_0} \right)$ where $\tilde{s}$ is independent of $N_c$.  If one takes the large $N_c$ limit prior to the large $s$ limit, and keeps only the leading behavior, one has
$\sigma^{\rm total} \le \frac{4 \pi}{m_\pi^2} \log^2 \left(N_c \right )$ up to corrections of relative order $1/\log(N_c)$.  Note that Eq.~(\ref{result}) satisfies this inequality by exactly a factor of $\frac{1}{2}$.  This factor of $ \frac{1}{2}$ is suggestive.  The derivation of the Froissart-Martin bound requires unitarity.  However, if one looks at the integral form of  Eq.~(\ref{pw2}) it is clear that  in the region of dominant contribution, the integrand at large $N_c$ is precisely $\frac{1}{2}$ of its unitarity bound (which occurs at $\delta^R=\pi/2$,  $\delta^I=0$).  Thus, the present result is natural in light of the Froissart-Martin bound.

To what extent is this result applicable to the physical world of $N_c$=3?  In the physical world, the total cross section for $\sqrt{s}$  well above $\Lambda{\rm QCD}$ is approximately independent of $s$\cite{PDG} as would be expected from the large $N_c$ analysis.   Over three orders of magnitude in $\sqrt{s}$, 
from $ 1.5 {\rm GeV}\; <\sqrt{s} < 1200 {\rm GeV}$, the cross section for proton-proton scattering varies by  only about 25\%.  Moreover, the  cross section is dominantly spin and isospin independent \cite{CohenGelman,NNexperimental} as predicted to occur at large $N_c$.  These results may suggest that the large $N_c$ analysis is of phenomenological relevance for the physical world of $N_c=3$.  However, this is not clear.  For example, above 1.5 GeV, the total cross section is predominantly inelastic; the elastic cross section is typically less than $\frac{1}{4}$  of the total cross section and by $\sqrt{s}$ of  several 10s of GeV, it drops to under 20\%.  While it has been argued that at truly asymptotically high energies \cite{BH} it approaches $\frac{1}{2}$, at large  $N_c$ this behavior is expected for all $s$ well above $\Lambda_{|rm QCD}$.    This implies that a substantial part of the cross section comes from regions where both the real and imaginary parts of  the phase shift are small.  This is at odds with the behavior expected at large $N_c$ where, as noted above, elastic scattering should be 50\% or higher.   Given this, it is perhaps not too surprising that the absolute prediction of Eq.~(\ref{result}) that $ \sigma^{\rm total} \approx 150 {\rm mb}$ is significantly larger than the empirical value of approximately 40 mb.  Ultimately, the reason that the large $N_c$ analysis for the magnitude of the total cross sections is not very predictive for the $N_c=3$ world is quite understandable.  Formally, one expects the  analysis to be predictive only when $\log (N_c) \gg 1 $.  Clearly this is not true for $N_c =3$. Whatever, the phenomenological significance for the world of $N_c=3$, the fact that at large $N_c$ the total cross section is calculable is, at the very least, of theoretical interest.

\begin{acknowledgments}
The author  thanks  Vojt\v{e}ch Krej\v{c}i\v{r}\'{i}k for helpful remarks.  The support of U.S. Department of Energy under grant DE-FG0293ER-40762 is gratefully acknowledged.

\end{acknowledgments}

\end{document}